\newcommand{\bfalpha}{\mbox{\boldmath $\alpha$}}

\newcommand{\bfTheta}{\mbox{\boldmath $\Theta$}}
\newcommand{\bfbeta}{\mbox{\boldmath $\beta$}}

\newcommand{\bfmu}{\mbox{\boldmath $\mu$}}

\newcommand{\bfphi}{\mbox{\boldmath $\phi$}}

\newcommand{\bfY}{\mbox{\boldmath $Y$}}
\newcommand{\bfy}{\mbox{\boldmath $y$}}

\newcommand{\bfx}{\mbox{\boldmath $x$}}

\newtheorem{theorem}{Theorem}

\documentclass[submit]{smj}

\Author{Pamela M. Chiroque-Solano\Affil{1}
}
\AuthorRunning{Pamela M. Chiroque-Solano}
  
\Affiliations{
\item  Centre for the Research and Technology of Agro-Environmental and Biological Sciences, Department of Mathematics, University of Trás-os-Montes e Alto Douro,
             5000-801 Vila Real, 
             Portugal.

}   

\CorrAddress{Pamela M. Chiroque-Solano, 
             University of Trás-os-Montes e Alto Douro,
             5000-801 Vila Real, 
             Portugal}
\CorrEmail{pchiroque@gmail.com}

\Title{Identifying Heterogeneity in Regression Compositional Data Integration with Many Categories}
\TitleRunning{Identifying heterogeneity in multiple CoDa}
\Abstract{
In compositional data, detecting which part of the whole delineates heterogeneity is important. The aim is to propose a procedure to quantify this term in the multivariate regression context without abandoning the data's natural restriction. A single probabilistic model with a hierarchical structure was built for multiple compositional data. An objective criterion based on skewness and kurtosis metrics provides support to characterize each component's performance as well as to assist in choosing one component as a reference avoiding model identifiability issues. The inference procedure was done under the Bayesian approach using the Hamiltonian Monte Carlo (HMC) method to obtain the posterior distribution of interest. The Kullback-Leibler divergence (KLD) from information theory and the Aitchison distance metrics are calculated to compute the similarity between compositions to compare scenarios in the model validation process. The proposal was motivated by a composition structure with high uncertainty in the Abrolhos Reefs of Brazil as a consequence of a dam rupture. The results support an understanding of patterns in the studied process recognizing local effects on each component as well as quantifying the precision parameter. These highlights contribute to characterizing the marine life community in areas that were affected by anthropogenic damage.
}
\begin{document}
\maketitle
\section{Introduction}
The role of environmental diversity in a community structure plays a crucial role factor in understanding an ecosystem in space and time \citep{Yang2015, Thompson2021}. Community structures delineate the interplay between a variety of species and habitat diversity following the principle that in a particular heterogeneous area, more species coexist \citep{Zeppilli2016, Heidrich2020}. Besides the changes due to the climate aspects, these ecosystems are continuously threatened by anthropogenic disturbances. Nowadays, advanced technology assists us in the data collection process. New data collection devices and sophisticated algorithms are representing a complex system 
more realistically \citep{Roelfsema2021}. 

As in other ecosystems in marine ecology, coral cover data sets allow us to recognize patterns related to the benthic coral reef community. Identifying and quantifying these effects represents reliable sources to characterize its dynamics. Understanding the marine ecosystem is essential to monitoring, evaluating, and managing the functioning of these communities. Similarly to \citep{Gross2015, Vercelloni2020}, this work focuses on modeling the benthic composition structure. In particular, it concentrates on identifying which part of the whole delineates heterogeneity when multiple compositional data (CoDa) sets are modeled simultaneously. It is advantageous to quantify this term in the multivariate regression context without abandoning the CoDa natural restriction.

Compositional Data is characterized by the case where each observation is a vector that occupies a restricted space where variables are non-negative and sum to one or any other given constant. It contains only relative information as the variables are parts of some whole. This relative information can be modeled as abundance data or relative abundance from count data \citep{Bacon2011}, and their construction allows the incorporation of temporal and spatial factors or other nuisance parameter nested with sites \cite{Allen2017, Chong2018}. Since the sample space for CoDa is radically different from the real euclidean space associated with unconstrained data,  classical multivariate statistical techniques should not be used directly to study it \cite{Aitchison1982, Pawlowsky-Glahn12006, Egozcue2011}. Many strategies such as data transformation were proposed to use traditional multivariate models on CoDa \cite{Aitchison1986, Boogaart2006}. For example, the additive log ratio and the multivariate Box-Cox transformations are available, but the results are only interpretable in the transformed space and have no straightforward meaning.
\subsection{Preliminaries}
An $n$ sized CoDa sample is given by a collection of vectors $\bfy_i=(y_{i1}, \ldots,y_{iC})'$ for  $i=1,\ldots,n$, where each component $y_{ic} \in (0,1)$ represents the proportion of component $c=1,\ldots,C$, for observation i. What is peculiar in this kind of data is that there is a relationship between components of the vector $y_i$. The associated random vector $\bfY_i$ is distributed in $C\geq 2$ dimensions subject to $\sum_{c=1}^{C}Y_{ic} = 1$. If a random vector $\bfY$ follows the Dirichlet distribution on the $C$-dimensional hyperplane or closed simplex $\mathcal{S}^{C}_{(1)}$ then this is denoted by $\bfY \sim \mathcal{D}_{C}(\bfalpha)$ on $\mathcal{S}^{C}_{(1)}$, where $\bfalpha$ is a $C$-dimensional vector of positive parameters and 
$\mathcal{S}^{C}_{(1)} = \left\{(Y_1,\ldots,Y_C)^{\prime}: y_c>0, 1\leq c \leq C, \sum_{c=1}^{C}Y_c = 1  \right\}$ or  $\mathcal{S}^{C}_{(1)} =   \{  \bfY: \bfY \in \mathbb{R}^{C}_{+}, \bf1_{C}\bfY = 1\}$, where $\mathbb{R}^{C}_{+}$ is n-dimensional positive orthant.

The expected value for each dimension $c$ of $\bfY$ is given by the contribution for the respective $\alpha_c$ parameter, that is, $E[Y_{c}] = \frac{\alpha_{c}}{\phi}$, where $\phi = \sum_{c=1}^{C}\alpha_{c}$. The variance is $Var(Y_{c}) = \frac{\alpha_{c}(\phi-\alpha_{c})}{\phi^2(\phi+1)}$ and the covariances $Cov(Y_c,Y_{c'}) = \frac{-\alpha_{c}\alpha_{c'}}{\phi^2(\phi+1)}$. The Dirichlet's probability density function is given by
\begin{eqnarray}
f(\bfY)&=& \frac{\Gamma(\alpha_{1}+\ldots+\alpha_{c})}{\prod _{c=1}^{C} \Gamma(\alpha_{c})} \prod _{c=1}^{C} Y_{c}^{ \alpha_{c}-1}, \quad a_c > 0, Y_c \in (0,1) , \sum_{c=1}^{C}Y_c=1.
\end{eqnarray}
Regression issues for CoDa were introduced at the end of the nineteen eighties by \cite{CampbellMosimann1987} as they proposed the Dirichlet covariate model. This approach was later widely studied by \cite{Hijazi2006, Gueorguieva2008, Hijazi2009}. They included the predictor term $\bfx^{\prime}_{i}\beta$ and defined the expected value as $E(Y_{c}\mid \mathbf x_i)$ proving satisfactory properties to understand the relationship between the composition with other variables. 

Over the last two decades, a wide variety of statistical models have been widely applied in many fields such as in microbiology \citep{Gloor2017, Tang2018, Espinoza2020}, market share by \cite{Morais2018}, geology by \cite{Barcelo1996, Pawlowsky-Glahn12006}, agriculture by \cite{Brewer2005}, psychiatric by \cite{Gueorguieva2008} and telecommunication and information theory by \cite{Thomas2006}. This brings forward the need to analyze various challenges such as the presence of outliers \citep{Barcelo1996} or assumptions like homoscedasticity. Another characteristic of some compositional data is the presence of null observed values in one or more categories. A Dirichlet regression model with observed zeros proposed by \cite{Tsagris2017} involves a computational-intensive effort. In the same context, \cite{Chong2018} implemented a multinomial regression approach to model changeable relative abundances due to the environmental gradients.

In the context of the Bayesian approach, \cite{Brewer2005} introduces a hierarchical model for compositional data and presented analyses of more complex situations via Markov chain Monte Carlo sampling. \cite{Minaya2021} used the Integrated Nested Laplace Approximation (INLA) to deal with such a regression model. Furthermore, \cite{Merwe2018} studied Bayesian fitting of Dirichlet Type I and II distributions. Additionally, \cite{Sean2018} developed a method for Bayesian regression modeling of compositional data including a new restriction in the mean of the distribution E[Yc], namely $\sum \mu_{ij} = 1$.

This proposal permits us a simple interpretation, extending the basic Dirichlet regression model proposed by \cite{Maier2014} in two ways. The first contribution deals with the model identifiability and the second with the integration of multiple CoDa through building a hierarchy \cite{McCullagh1989}. This proposal induces a flexible structure to detect and quantify the presence of heteroscedasticity, skewness, kurtosis and outlier occurrence. This paper uses the Bayesian approach to make inference on the multilevel or hierarchical structure of the model, evidencing the borrowed strength among multiple CoDa sets through a common mean.

Following this introduction, the Dirichlet regression model is characterized by an alternative parameterization to describe the Dirichlet density in
terms of the variate mean and a precision parameter. As a result, this parameterization induces a constraint in the model delineation. An objective metric is recommended to assist in the choice of the reference component in the implementation process. The procedure to lead with the reference component choice is based on the shape of the distribution. It is presented in Section \ref{sec:reference}. The multilevel Dirichlet Regression model is presented in Section \ref{sec:model}. Section \ref{sec:inference} focuses on the inference procedure based on the Bayesian approach. In addition, adequate metrics that helped in the model evaluation process are presented.
In Section \ref{sec:simulation} the process of data simulation is exemplified by considering scenarios with high, low and medium variability.
The application in Section \ref{sec:application} presents the performance of the proposed model on a marine ecology data set. 
The conclusion and future works, advantages and limitations of the proposed methodology are discussed in Section \ref{sec:conclusion}.
\section{Dirichlet parametrization}\label{sec:reference}
The common parametrization in \cite{Hijazi2009}, is focused on the mean of the empirical process. The alternative parametrization in \cite{Maier2014} allows us to consider heteroscedastic information. The idea behind this alternative parametrization generalizes was developed in the univariate beta regression \citep{Cribari2010}. They considered a term to describe the distribution’s precision as the sum of both beta distribution's parameters. This is extended to modeling the precision term $\phi=\sum_{c=1}^{C}\alpha_{c}$, describing the parameter vector of the Dirichlet regression as $\alpha_{c}=\mu_{ic}\phi$, where $\mu_{ic} = E(Y_{c}\mid \mathbf x_i)$ represents the level term, where $\bfx_i$ are the covariates. This implies a model with $C+1$ parameters to estimate, where $C$ are unrestricted. 

In other words, the model suffers an identification problem because the $\bfalpha$ vector generates a unique $\phi$. Identifiability issues are common in CoDa modeling and to avoid this situation a valid strategy is to choose a reference component $c^*$ whose effect parameters are not estimated, while the others are parameterized with respect to it. This approach guarantees that the $\phi$ term quantifies the heterogeneity of the process. The theoretical results are valid no matter which category is chosen, but in practice, some choices can lead to better results interpretation and more stability concerning the experiment itself. A practical guide to choosing this reference component was not found in the literature, the usual selection being arbitrarily the first or last $C$ component.
\subsection{Objective reference component}\label{sec:proposal}
A reference component should not be one of scientific interest. In Statistics, information is gathered from variability. A reference component is expected, then, to be little prone to outliers and to have little variability, as such it should not add much to the process’s overall accumulated knowledge. To put it more technically, a reference component is recommended to be chosen by having small values of asymmetry and kurtosis. This configures it as being well-behaved, which is close to Normal performance. 

The procedure to choose a steady component consists of fitting a standard, non-regression Dirichlet model to the data. The stochastic representation of Theorem \eqref{theorem1} presented in  \citep[chapter 2]{Wang2011}, allows us to use a sequence of independent gamma variables as an alternative representation for the Dirichlet $(y_1, \cdot, y_C)\prime$ random vector.
\begin{theorem}\label{theorem1}
	$	\bfy = (y_1, \ldots,y_C)^\prime \sim \mathcal{D}_{C}(\bfalpha)$ on $\mathbb{T}_C$ iff
	$y_i \stackrel{\text{d}}{=} \frac{w_c}{w_1+\ldots+w_C}, c= 1, \ldots, C,$
	where $w_c \sim \mathcal{G}(\alpha_c,1)$ and $\{w\}^{n}_{c=1}$ are mutually independent.
\end{theorem}
This theorem can be interpreted as disassembling the composition through the fitted $\bfalpha$ vector parameters. This will help to calculate the kurtosis $(3+6/\alpha)$ and central symmetry (skewness $2/\sqrt{\alpha}$) metrics related to the Gamma distribution from each component.

The decision criterion considers the lowest values of kurtosis described by high implying a platykurtic distribution, wider peak around the mean and thinner tails \citep{Westfall2014}. It is an indicator that a variable has fewer outliers. As a summary a recommendation for the reference component is to consider $c^*$ whose $\alpha_{c^*}$ value is the highest one whereas the basic Dirichlet distribution was fitted. This guarantees that the $c^*$ component has low skewness and kurtosis.

To illustrate this behavior the skewness and kurtosis statistics are to be calculated for each component independently after the procedure that disassembles the Dirichlet vector into Gamma random variables. Due to the valuable information added, the component with the lowest metrics is to be chosen as a reference component. Next, a simulation example is presented to illustrate the dynamic.
\subsubsection{Artificial choices illustration}
\begin{figure}[!ht]
	\centering
	\includegraphics[width=1\linewidth]{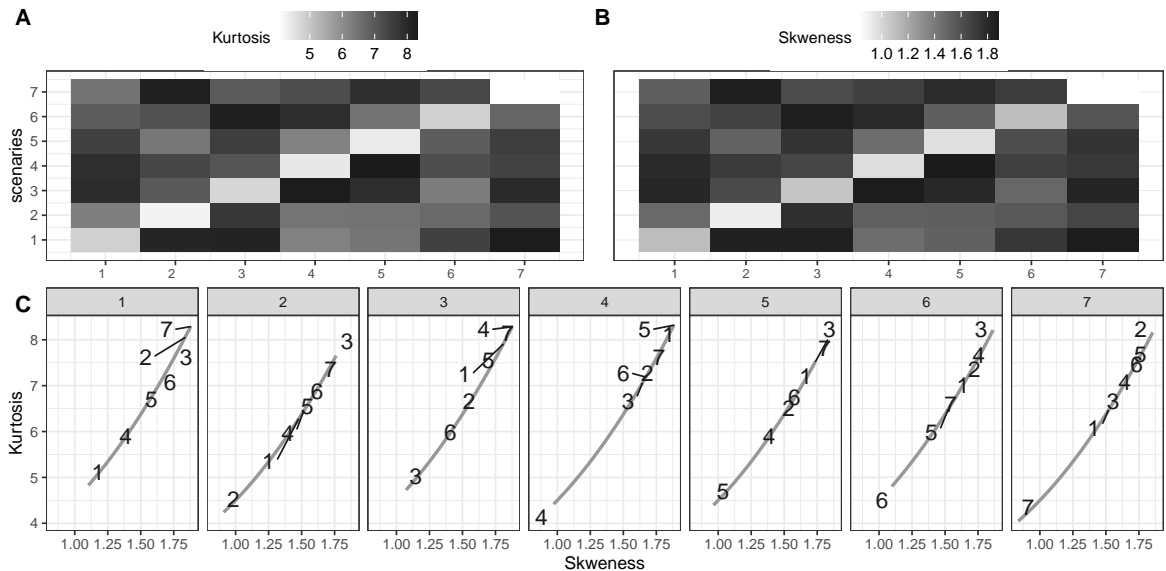}
	\caption{Seven scenarios, averaging over 100 replicates. (A) kurtosis and (B) skewness resulting from the stochastic gamma representation applying Theorem \ref{theorem1}. (C) Representation of the relationship between kurtosis and skewness.}
	\label{fig:repstocSim7}
\end{figure}

The parameter vector $\bfalpha$ with components $\alpha_c , c = 1, \cdots , C = 7$ was sampled following a uniform distribution with parameters varying from $1.1$ to $1.9$, $\mathcal{U}(1.1,1.9)$. Seven artificial scenarios were built to exemplify this choice. Each of the seven components was chosen as the reference for one scenario. For each reference, the respective $\alpha_c$ value was added by a random variable with normal distribution and mean 4 and variance 1, that is, $\mathcal{N}(4,1)$. This was to ensure that the chosen reference had the largest $\alpha_{c}$ value. This characterization allows us to describe equivalent probability distributions for all scenarios. Additionally, estimated values of the $\alpha_{c}$ parameters will be the same (on average) in different positions of the seven-dimensional vector. This exercise was repeated $100$ times for each scenario. The $\alpha$ values were estimated and the kurtosis and skewness metrics were calculated. Table \ref{Table:Artificial7} contains the random seed used in the artificial exercise for the seven scenarios.
\begin{table}[h!]
\scriptsize
\begin{minipage}[t]{0.55\textwidth}
\begin{tabular}{rrrrrrr}
\hline
Sc & C & $\alpha_c$ & $\phi$ & E & Skew & Kurt\\ 
		\hline
		\multirow{7}{*}{1}&
		1 & 4.59 & 13.76 & -7.99 & 0.93 & 4.31 \\ 
		&   2 & 1.88 & 13.76 & -7.99 & 1.46 & 6.19 \\ 
		&   3 & 1.87 & 13.76 & -7.99 & 1.46 & 6.20 \\ 
		&   4 & 1.13 & 13.76 & -7.99 & 1.88 & 8.29 \\ 
		&   5 & 1.26 & 13.76 & -7.99 & 1.78 & 7.74 \\ 
		&   6 & 1.36 & 13.76 & -7.99 & 1.71 & 7.40 \\ 
		&   7 & 1.65 & 13.76 & -7.99 & 1.56 & 6.63 \\ 
		\multirow{7}{*}{2}& 
		1 & 1.63 & 14.04 & -8.20 & 1.57 & 6.68 \\ 
		&   2 & 5.27 & 14.04 & -8.20 & 0.87 & 4.14 \\ 
		&   3 & 1.57 & 14.04 & -8.20 & 1.60 & 6.83 \\ 
		&   4 & 1.78 & 14.04 & -8.20 & 1.50 & 6.37 \\ 
		&   5 & 1.25 & 14.04 & -8.20 & 1.79 & 7.78 \\ 
		&   6 & 1.26 & 14.04 & -8.20 & 1.78 & 7.76 \\ 
		&   7 & 1.27 & 14.04 & -8.20 & 1.78 & 7.73 \\ 
		\multirow{7}{*}{3}& 
		1 & 1.55 & 13.72 & -7.74 & 1.61 & 6.88 \\ 
		&   2 & 1.44 & 13.72 & -7.74 & 1.67 & 7.17 \\ 
		&   3 & 4.17 & 13.72 & -7.74 & 0.98 & 4.44 \\ 
		&   4 & 1.81 & 13.72 & -7.74 & 1.49 & 6.31 \\ 
		&   5 & 1.72 & 13.72 & -7.74 & 1.52 & 6.48 \\ 
		&   6 & 1.32 & 13.72 & -7.74 & 1.74 & 7.54 \\ 
		&   7 & 1.71 & 13.72 & -7.74 & 1.53 & 6.52 \\ 
		\multirow{7}{*}{4}& 
		1 & 1.17 & 13.32 & -8.00 & 1.85 & 8.13 \\ 
		&   2 & 1.14 & 13.32 & -8.00 & 1.87 & 8.25 \\ 
		&   3 & 1.89 & 13.32 & -8.00 & 1.46 & 6.18 \\ 
		&   4 & 4.59 & 13.32 & -8.00 & 0.93 & 4.31 \\ 
		&   5 & 1.66 & 13.32 & -8.00 & 1.55 & 6.62 \\ 
		&   6 & 1.30 & 13.32 & -8.00 & 1.75 & 7.60 \\ 
		&   7 & 1.57 & 13.32 & -8.00 & 1.59 & 6.81 \\ 
		\hline
	\end{tabular}
\end{minipage}
\begin{minipage}[t]{0.5\textwidth}
	\begin{tabular}{rrrrrrr}
		\hline
		Sc & C & $\alpha_c$ & $\phi$ & E & Skew & Kurt\\ 
		\hline
		\multirow{7}{*}{5}& 
		1 & 1.81 & 12.50 & -7.38 & 1.49 & 6.31 \\ 
		&   2 & 1.31 & 12.50 & -7.38 & 1.75 & 7.58 \\ 
		&   3 & 1.85 & 12.50 & -7.38 & 1.47 & 6.24 \\ 
		&   4 & 1.74 & 12.50 & -7.38 & 1.52 & 6.45 \\ 
		&   5 & 2.93 & 12.50 & -7.38 & 1.17 & 5.05 \\ 
		&   6 & 1.70 & 12.50 & -7.38 & 1.53 & 6.52 \\ 
		&   7 & 1.16 & 12.50 & -7.38 & 1.86 & 8.17 \\ 
		\multirow{7}{*}{6}& 
		1 & 1.71 & 14.11 & -7.71 & 1.53 & 6.52 \\ 
		&   2 & 1.72 & 14.11 & -7.71 & 1.53 & 6.49 \\ 
		&   3 & 1.47 & 14.11 & -7.71 & 1.65 & 7.10 \\ 
		&   4 & 1.80 & 14.11 & -7.71 & 1.49 & 6.33 \\ 
		&   5 & 1.74 & 14.11 & -7.71 & 1.52 & 6.45 \\ 
		&   6 & 4.14 & 14.11 & -7.71 & 0.98 & 4.45 \\ 
		&   7 & 1.54 & 14.11 & -7.71 & 1.61 & 6.90 \\ 
		\multirow{7}{*}{7}& 
		1 & 1.58 & 14.29 & -8.02 & 1.59 & 6.79 \\ 
		&   2 & 1.63 & 14.29 & -8.02 & 1.57 & 6.69 \\ 
		&   3 & 1.65 & 14.29 & -8.02 & 1.56 & 6.63 \\ 
		&   4 & 1.61 & 14.29 & -8.02 & 1.58 & 6.73 \\ 
		&   5 & 1.77 & 14.29 & -8.02 & 1.50 & 6.39 \\ 
		&   6 & 1.16 & 14.29 & -8.02 & 1.86 & 8.17 \\ 
		&   7 & 4.89 & 14.29 & -8.02 & 0.90 & 4.23 \\ 
		\hline
	\end{tabular}
\end{minipage}
\caption{True values for seven artificial scenarios (Sc), and seven-dimensions, C=7, vector $\bfalpha$, $\phi = \sum_{c=1}^{c=7}\alpha_c$, kurtosis $3+6/\alpha$, skewness $2/\sqrt{ \alpha}$ and E: entropy metric $H(y) = \log\left(\frac{\prod^{C}_{c=1}\Gamma(\alpha_c)}{\Gamma(\phi)}\right) + (\phi-C)\digamma(\phi) - \sum_{s=1}^{C}(\alpha_s-1)\digamma(\alpha_s)$ where $\phi = \sum_{c=1}^{C}$, $\digamma$ is a digamma function and $C$ number of components. }\label{Table:Artificial7}
\end{table}
A graphical representation is presented in Figure \ref{fig:repstocSim7} where $\phi$ varies between (12.5,14.11), the lowest values for kurtosis and skewness are in the diagonal (Figure \ref{fig:repstocSim7} A, B) for each scenario. Note that, the nonlinear relationship between kurtosis and skewness is evidenced (Figure \ref{fig:repstocSim7} C), as expected in the gamma family \citep{Westfall2014}. This illustration evidences the relationship between the highest alpha parameter component with the lower skewness and kurtosis values. That means that the precision estimation is maintained regardless of which component is chosen if it is related to the high alpha and it achieves the lowest skewness and kurtosis. This decision procedure provides us with a practice method of how to choose the reference component in a straightforward manner.
\section{Hierarchical Dirichlet regression under the Bayesian approach}\label{sec:model}
The basic Dirichlet regression model, originally proposed by \cite{Maier2014}, was worked by \cite{Holger2018} using Markov chain Monte Carlo (MCMC) methods for sampling from the posterior distribution. 

Following the notation in \cite{Maier2014} a multilevel structure through the integration of the $l=1,\ldots,L$ datasets including a P$-$dimensional vector of regressors $x$ is proposed. The distribution’s precision is considered as well. The effects are quantified by the coefficient vector $\beta_{cl}$ relative to component $c$ and the dataset $l$ and once a reference is chosen the effects related to it are set as zero, $\beta_{c^*} = 0$. The linear predictors for the expected value for observation $Y_{cl}$ is denoted as $\mu_{icl} = E(Y_{icl}\mid \mathbf x_i)$ and the precision part $h(\phi_{il}) = \mathbf{z}_i^{\prime}\theta_l$ are given by
\begin{eqnarray}
	\mu_{icl} &=& E(Y_{icl}\mid \mathbf x_i) = \frac{\exp( \mathbf x_i^{\prime}\mathbf\beta_{cl})}{\sum_{d=1}^{C}\exp(\mathbf x_i^{\prime}\mathbf\beta_{dl})} \in (0,1) \\ \nonumber
	\phi_{il} &=& \exp(\mathbf z_i^{\prime}\theta_{l}), \quad \theta_{l} \in \mathbb{R} \nonumber \\
	\beta_{cl} &=& \beta_{c}+ \epsilon_{\beta_{l}} \nonumber\\
	\theta_{l} &=& \theta+ \epsilon_{\theta_{l}}\nonumber
\end{eqnarray}\label{eq:mu-phi}
The $z$ terms are drivers that can contain the same information as $x$ but can contain other variables that impact only the distribution’s precision. The $\epsilon$ terms encode the hierarchy effects. The $\bfbeta$ and $\theta$ parameters capture the dataset's information sharing. The expectation vector $\mu_{il{c^*}}$  for $c^*$ is given by $\mu_{ilc} = \frac{1}{\sum_{d=1}^{C}\exp(\mathbf x_i^{\prime}\mathbf\beta_{dl})}.$

Define the parameters $\alpha_{ilc} = \mu_{ilc}\phi_{l}$ to return to the Dirichlet distribution’s original $\alpha_{ilc}$ parameters with $\phi_l = \sum_{c=1}^{C}\alpha_{cl}$. The Dirichlet distribution with
regressed mean is denoted by $Y_{il}\mid \bfx_i \sim \mathcal{D}(\bfmu_{il},\phi_l), \quad i=1,\ldots,n, \quad l=1,\ldots,L$. We obtain the density function associated with our proposed model
\begin{eqnarray}
f(\bfy_{il}\mid  \mu_{il},\phi_l) = \frac{\Gamma(\sum_{c=1}^{C} \mu_{icl}\phi_{l})}{\prod _{c=1}^{C} \Gamma(\mu_{icl}\phi_{l})} \prod _{c=1}^{C} y_{icl}^{ \mu_{icl}\phi_l - 1}, \quad \mu_{icl}\in(0,1), \phi_{l} >0, \sum_{c=1}^{C}y_{icl}=1.
\end{eqnarray}
\subsection{Inference: Posterior sampling}\label{sec:inference}
Let $\Theta =(\bfbeta,\bfphi)$ be the vector of parameters and the likelihood function $L(\bfTheta \mid {\bfy} )=\prod_{i=1}^{n}\prod_{l=1}^{L}f(\bfy_{il} \mid \bfTheta)$. More specifically, 
\begin{eqnarray}
\log L(\bfTheta \mid \bfy ) = \sum_{i=1}^{n}\sum_{l=1}^{L}( \log \Gamma(\sum_{c=1}^{C} \mu_{icl}\phi_{l})-  \sum_{c=1}^{C}\log \Gamma(\mu_{icl}\phi_l)+  \sum_{c=1}^{C}(\mu_{icl}\phi_l-1)\log(y_{icl})),
\end{eqnarray}
\noindent where the $\mu_{icl}$ are functions of $\bfbeta_{cl}$ which are unknown components to be estimated. The vector $\bfy_l=(\bfy_{1l},\ldots,\bfy_{nl})$ denotes all the information available provided by the data in location $l$, where each $y_{1l}$ is a C$-$dimensional vector.
To complete the model specification we assigned an appropriate proper prior distribution to the parametric vector $\bfTheta$. The prior distribution is chosen so that there is little prior information. Prior independence between the parameters is assumed. The choice of the proper independent prior distributions was driven by the choice of making inference with minimum subjective prior information about parameter correlation. The parameters $\theta_l = \log(\phi_l)$ and $\beta_{cl}, c= 1, \cdot, C$ and $l = 1, \cdot, L$ are normally distributed with zero mean and unknown precision. A prior for the precision is set as a half-Cauchy.
The posterior distribution is given by
\begin{eqnarray}\label{eq:posterior}
\pi(\bfTheta \mid \bfy) \propto L(\bfTheta\mid \bfy)\prod_{l}^{L}\pi(\phi_l)\prod_{c}^{C}\pi(\beta_{cl})
\end{eqnarray}
Since the joint posterior distribution in (\ref{eq:posterior}) does not have a known closed form, we propose the use of MCMC methods to obtain samples from it. The inference procedure was done under the Bayesian approach using the Hamiltonian Monte Carlo (HMC) method to obtain approximations of the posterior marginal distributions of interest. Sampling from the distribution whose density is in equation (\ref{eq:posterior}) is done using the No-U-Turn-Sampler algorithm implemented in the Stan software.
\subsection{Diagnostic metrics}\label{sec:diagnostic}
It is necessary to establish comparison metrics to measure model fit in accordance with the nature of the data to guarantee a good fitting and predictive performance. To achieve this the Aitchison distance and the Kullback-Leibler divergence describe the capability of the model to recover the composition structure. The coverage and Root Mean Squared Error (rMSE) deal with the precision and accuracy of the estimates of the parameters. The deviance information criterion (DIC) \cite{Spiegelhalter2002} and the widely applicable information criterion (WAIC) \cite{Gelman2014} statistics assess the model fit.

Goodness-of-fit metrics compare the observed values with their fitted values. Let $\bfY_1=(y_{11},\ldots,y_{1C}) \in \mathcal{S}_{(1)}^{C}$ be the observed values and $\bfY_2=(y_{21},\ldots,y_{2C}) \in \mathcal{S}_{(1)}^{C}$ the fitted ones. Then the Aitchison's Distance \cite{Aitchison1986} is define by $\Delta(\bfY_1,\bfY_2) = \sqrt{\sum_{i=1}^{n}(ln(r_{i}) - D)^2}$ , where $r_{i} = \frac{y_{1i}}{y_{2i}}$ and  $D = \frac{\sum_{i=1}^{C}ln(r_{i})}{C}$. Based on the divergence between two Dirichlet distributions $D(f(\bfy)\|g(\bfy))$, it can be useful to express the relative entropy (a.k.a. Kullback-Leibler divergence) between these two compositions as $KL(\bfy_1,\bfy_2) = \sum_{c=1}^{C}y_{1c} \log \left( \frac{y_{1c} }{y_{2c}}\right)$ \citep{Kullback1997,Thomas2006, Vidal2007, Ionas2021}. The factor $ \log(\frac{y_{1c}}{y_{2c}})$ can be interpreted as the information gained in predicting the event related to component $c$ where lower values imply a better fit.
A metric to measure the predictive accuracy is the coverage which is the proportion of a test set whose observations $\bfy_{i}$ fall inside some interval of their predictive distribution $\mathcal{P}_i$ \cite{GelmanBDA3}. 

The 95\% coverage is given by $\mathrm{C}^{(95\%)} = \frac{1}{n} \sum_{i=1}^{n} \left[ \mathcal{I} \left( \mathcal{P}_i^{(0.025)} < y_i <  \mathcal{P}_i^{(0.975)}  \right)\right],$ where $\mathcal{I}$ is the indicator function and $\mathcal{P}^{(q)}$ is the estimated $q^{th}$ quantile of $\mathcal{P}_i$.
Another useful metric is the root Mean Squared Error (rMSE). It can be used to quantify prediction quality, but it is also useful in simulation exercises to measure the adequacy of parameters estimation. It is calculated as $rMSE=100 \sqrt{\frac{1}{J}\sum_{j=1}^{J}\left(\widehat{\mbox{estimated}}- \mbox{true}\right)^2}$.
A model is preferred when it has coverage close to 95\% and the smallest $rMSE$, $KL$ divergence, and Aitchison’s Distance.

The deviance information criterion (DIC) \citep{Spiegelhalter2002} is defined by $\mathrm{DIC}=D\left(\bar{\mathbf{\Theta}}\right)+2p_D$, where the deviance  $D\left(\mathbf{\Theta}\right)=-2\log{\left(L\left(\mathbf{\Theta}\mid y\right)\right)}$ and $y$ denotes all information provided by the data and $\mathbf{\Theta}$ is the vector of parameters and $L\left(\mathbf{\Theta}\mid y\right)$ is the likelihood function.  The term $p_D=\overline{D\left(\mathbf{\Theta}\right)}-D\left(\bar{\mathbf{\Theta}}\right)$, contains $\overline{D\left(\mathbf{\Theta}\right)}$ and $\bar{\mathbf{\Theta}}$ which denote the posterior mean of $D\left(\mathbf{\Theta}\right)$ and $\mathbf{\Theta}$, respectively. The widely applicable information criterion (WAIC) statistic was calculated using $\hat{\mathrm{elppd}}_{WAIC} = \mathrm{lppd} - p_{WAIC},$ where $\mathrm{lppd} = \log \prod_{i=1}^{n} p_{post} (y_i) = \sum_{i=1}^{n} \log\int p(y_i \mid \theta) p(\theta\mid y)d\theta$ is the $\log$ pointwise predictive density. $\mathrm{lppd}$ is computed using $\sum_{i=1}^{n} \log \left(\frac{1}{L} \sum_{l=1}^{L} p(y_{i} \mid \theta_{l}) \right)$ and the correction term $p_{WAIC} = \sum_{i=1}^{n} V_{j=1}^{J} (\log p(y_i \mid \theta_{j}))$, where $V_{j=1}^{J}$ represents the sample variance, $V_{j=1}^{J}a_j = \frac{1}{J-1}\sum_{j=1}^{J}(a_s - \hat{a})^2$. Thus we use $\mathrm{WAIC} = -2\hat{\mathrm{elppd}}_{WAIC} $. Lower values of DIC and WAIC imply higher predictive accuracy \citep{Gelman2014}.
\subsection{Direct interpretation of the $\beta$ effect on each component}
The effects of predictors $\beta$ are estimated on the composition of $C-1$ dimensions and the log-ratios should be used to interpret these effects. Negative effects values of a predictor on a log ratio between specific and reference component mean that the coverage of the specific component usually decreases after the predictor’s influence.
\section{Simulation study}\label{sec:simulation}
In order to understand the behavior of variability and information quantification of the proposal a simulation study was performed. It was motivated by the relationship between the precision and entropy values (Table \ref{Table:Artificial7}). 

\begin{figure}[!ht]
	\centering
	\includegraphics[width=1\linewidth]{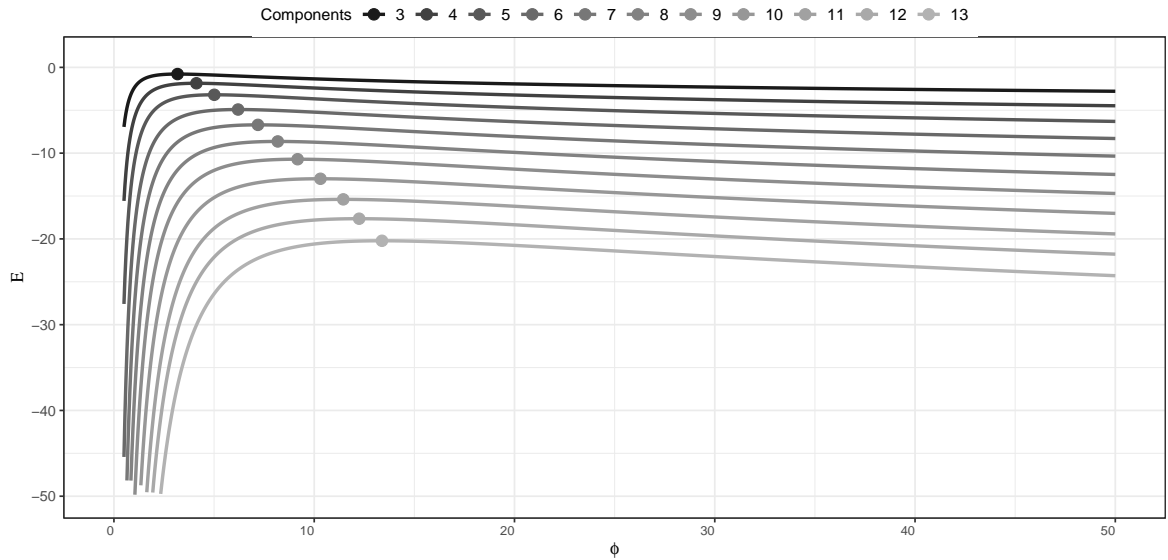}
	\caption{Relationship between entropy and $\phi = \sum_{c=1}^{C}\alpha_c$ for eleven $c-$ dimensional scenarios. The solid circle indicates the point where each curve achieves the maximum entropy.}
	\label{fig:EP}
\end{figure}

The simulation experiment was structured to evaluate different precision values chosen conditionally to the entropy metric. The exercise consists of eleven cases with different $c= 3, \cdots ,13$ dimensional vectors which were sampled revealing the nonlinear relationship between entropy, defined below, and $\phi = \sum_{c=1}^{C}\alpha_c$ (Figure \ref{fig:EP}). The entropy metric for the Dirichlet distribution is defined as $H(y) = \log\left(\frac{\prod^{C}_{c=1}\Gamma(\alpha_c)}{\Gamma(\phi)}\right) + (\phi-C)\digamma(\phi) - \sum_{s=1}^{C}(\alpha_s-1)\digamma(\alpha_s)$,
where $F$ is the digamma function. Figure \ref{fig:EP} represents this relationship.
For simplicity, the simulation does not consider a regression structure for $\phi$. Instead, for simulating the multiple dataset scenario, its value changes slightly around a predetermined arbitration. Based on the previous exercise, the simulation study used three precision values, 13, 5 and 2, describing high, medium and low entropy values, respectively, -1.5584, -0.9009 and -0.9488. The number of components was chosen as $C=3$.
The used covariate is a categorical value that represents the origin of the dataset. Therefore, the effects quantify the difference between the l = 4 datasets. They are chosen as small variations around $\beta_1 = 2, \beta_2 = 3, \beta_3 = 0$. These variations (random effects) are namely $0.002, 0.003, -0.002$ and $-0.003$.
Equations in \eqref{eq:mu-phi} were used to calculate the parameter vector $\bfalpha$ where $\alpha_{c}=\mu_{ic}\phi$ with which $N$ observations from the Dirichlet distribution were sampled for each dataset l. One hundred replications were performed for $N$ = 10, 15 and 30 for each value of $\phi$.
The resulting data for each replication was used to fit a model using the aforementioned prior.
\subsection{Computational Time}\label{sec:comp-time}
To obtain samples of the posterior distribution, the MCMC algorithm was implemented using the Stan software \citep{StanManual2018}, and two modeling strategies were implemented. One of them reduced the mean computational time with respect to the other passing from 100.859 to 54.1941 seconds for each chain. The slower one, called Alg1-pgr2, was sampled from the posterior distribution using arrays. The other, Alg2-pgr2G, was implemented using vectorization. The procedures described in Section \ref{sec:inference} were implemented to monitor the chains’ convergence.
\begin{table}[h!]
\centering\footnotesize
\begin{tabular*}{\textwidth}{ c@{\extracolsep{\fill}}c@{\extracolsep{\fill}}c@{\extracolsep{\fill}}c@{\extracolsep{\fill}}c@{\extracolsep{\fill}}c@{\extracolsep{\fill}}c@{\extracolsep{\fill}}c@{\extracolsep{\fill}}c@{\extracolsep{\fill}}c@{\extracolsep{\fill}}}
		\toprule
		\multirow{2}{*}{True $\phi$}&\multirow{2}{*}{$N$} & \multirow{2}{*}{Algorithm} & \multicolumn{2}{c}{Parameters fitted} &\multicolumn{4}{c}{Prediction performance} \\ 	
		\cline{4-5}\cline{6-9}
		& &  & Cover 95\% & rMSE \% & aDist$^\mathcal{P}$ & Cover$^\mathcal{P}$ 95\% & rMSE$^\mathcal{P}$ \%& KL$^\mathcal{P}$ \\ 
		\midrule
		\multirow{6}{*}{13}
		&\multirow{2}{*}{10} & pgr2 & 0.958 & 0.310 & 1.711 & 0.958 & 0.074 & 0.061 \\ 
		&                    & pgr2G & 0.958 & 0.297 & 1.707 & 0.977 & 0.073 & 0.060\\ 
		&\multirow{2}{*}{15} & pgr2 & 0.968 & 0.245 & 1.669 & 0.949 & 0.076 & 0.062\\ 
		&                    & pgr2G & 0.948 & 0.241 & 1.663 & 0.965 & 0.075 & 0.061 \\ 
		&\multirow{2}{*}{30} & pgr2 & 0.978 & 0.170 & 1.670 & 0.941 & 0.075 & 0.062 \\ 
		&                    & pgr2G & 0.947 & 0.167 & 1.670 & 0.956 & 0.074 & 0.062 \\
		\midrule
		\multirow{6}{*}{5}
		&\multirow{2}{*}{10} & pgr2  & 0.952 & 0.331 & 3.998 & 0.953 & 0.113 & 0.271 \\ 
		&                    & pgr2G & 0.948 & 0.351 & 4.010 & 0.976 & 0.113 & 0.276 \\ 
		&\multirow{2}{*}{15} & pgr2  & 0.967 & 0.265 & 4.050 & 0.948 & 0.113 & 0.270 \\ 
		&                    & pgr2G & 0.950 & 0.269 & 4.051 & 0.967 & 0.113 & 0.268 \\ 
		&\multirow{2}{*}{30} & pgr2  & 0.973 & 0.203 & 3.967 & 0.939 & 0.115 & 0.266 \\ 
		&                    & pgr2G & 0.946 & 0.195 & 3.968 & 0.957 & 0.114 & 0.263 \\ 
		\midrule
		\multirow{6}{*}{2}
		&\multirow{2}{*}{10} & pgr2  & 0.969 & 0.309 & 10.142 & 0.949 & 0.158 & 0.244 \\ 
		&                    & pgr2G & 0.938 & 0.369 & 10.241 & 0.978 & 0.161 & 0.242 \\ 
		&\multirow{2}{*}{15} & pgr2  & 0.978 & 0.251 & 10.121 & 0.947 & 0.160 & 0.248 \\ 
		&                    & pgr2G & 0.938 & 0.290 & 10.190 & 0.972 & 0.162 & 0.246 \\ 
		&\multirow{2}{*}{30} & pgr2  & 0.946 & 0.200 & 10.129 & 0.938 & 0.165 & 0.260 \\ 
		&                    & pgr2G & 0.934 & 0.217 & 10.159 & 0.960 & 0.165 & 0.256\\ 
		\bottomrule
	\end{tabular*}
	\caption{Evaluating fitted and prediction performance based on the 100 simulated datasets. $\phi$ indicates the precision, $N$ the sample size and the algorithm used to fit. Cover 95\%: 95\% coverage. rMSE: Root of the Mean-Squared Error. aDist: Aitchison distance. KL: Kullback Leibler divergence.}\label{table:simr}
\end{table}
\subsection{Parameters fitted} The parameter estimation accuracy is obtained through the rMSE with values between 0.2-0.3\%. For the highest precision value scenario ($\phi$= 13), the lowest rMSE value is related to the sample size $N=30$. Similar rMSE values were obtained when $\phi=5$ and $\phi=2$ for all $N$ sample sizes. On the other hand, on average all values of coverage show how well the model does in creating posterior predictive distributions that capture the true value. The scenario with $\phi=13$ showed the most stable coverage values. Although the results did not present high coverage probability they were approximately equivalent varying between 94-95\%. This is a good precision estimation result because the high coverage value is a result of the high-variance posterior predictive distribution. Coverage values around 95\% are preferred.
Thus, the proposed model helps us to predict values from this process.  
\subsection{Predictive performance} Predictions related to the lower values of $\phi=2,5$ present highest aDist, KL and rMSE. The coverage value remains around 95\% for all the sample sizes. The Kullback-Leibler divergence and Aitchison’s Distance are adequate metrics for compositional data. For $\phi=13$ the average information gain $KL(y_{obs},y_P)$ is the lowest for all sample sizes, as expected. The parameterization of the model using the precision parameter $\phi$ is particularly useful as a direct relation with data variability and the KL metric can be drawn.
On the other hand, Aitchison's metrics were extremely different when $\phi=5$ and $\phi=2$, approximately 4 and 10 respectively. The presence of $\phi$ also presents a direct expectation for the results in this case. The distances are notably smaller as $\phi$ grows.
\section{Application: Abrolhos bank}\label{sec:application}
The reef community composition dataset is captured through images that were processed following a semi-automatically approach which includes specialized algorithms such as deep neural networks in the procedure CoralNet platform \citep{Beijbom2015}. The relative cover was estimated from the identification of benthic organisms which were identified at nine broad taxonomic or functional groups describing the benthic composition community. The benthic community structure is represented as a composition vector with 19909 observations from the five more representative Abrolhos sites between 2006 and 2018 \citep{Carol2021}.

This work used the components of reef structures data of the Abrolhos bank, which are built by categories such as corals, fire coral, sponge, bryozoans, others, cca, cyano-bacteria, macroalgae and turf. Each one of these categories describes the benthic community and they can be expressed as proportions of a whole. The cover values of these components were jointly studied for all sites, described below, to understand their distinct pattern and variability.

The sessile benthic cover was sampled during austral summers between 2006 and 2018, using 100 fixed photo-quadrats (0.7 $m^2$ each) per year. Data was monitored in three inshore sites; Pedra de Leste (PLES), Sebastião Gomes (SGOM), Timbebas (TIMB) and three offshore sites, PAB2, PAB3 and PAB5; within the Parcel dos Abrolhos reef. Details related to the experimental design and the region can be found in \cite{Carol2021}.  All sampling has been done by biologist scuba divers using the photo quadrat technique, which were then processed by the CORALNET software. All the statistical analyses were conducted using R \citep{RR}.

The main objective of this application is to determine the heterogeneity of the multivariate coral reef compositional response in different sites evolving from different geographical patterns in the Abrolhos bank. To achieve that we model the variability effects by sites including a hierarchical structure. The first step is to obtain the reference component.

To perform the fit of the model we used the observation vector $y_{ilh}$ with nine components within each dataset (location) $l = 1, \cdots , L=6$. The model's rationale was defined via the relevance of compositional data described in Section \ref{sec:reference}. The model described in Section \ref{sec:model} was implemented in a Bayesian framework.

To contrast the impact of the habitat factor on the cover components, the information about the two habitats, namely the top and wall, was included as a predictor in the model. There were a total of $n=19909$ observations. 
\begin{figure}[!ht]
	\centering
	\includegraphics[width=1\linewidth]{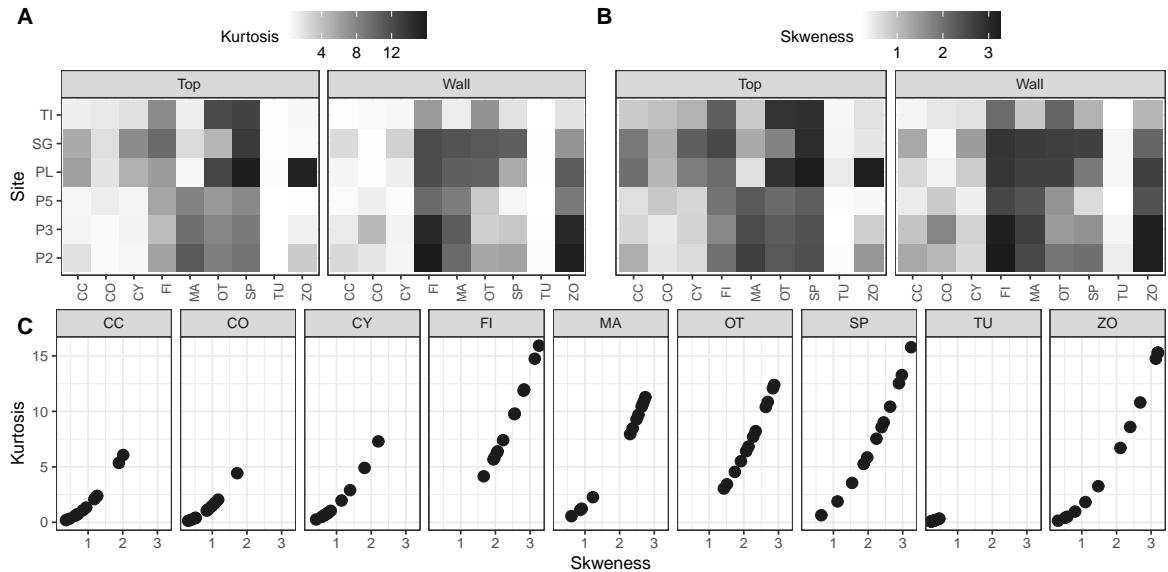}
	\caption{Kurtosis and skewness metrics on values resulting from the stochastic gamma representation by site and habitat}
	\label{fig:repstoc}
\end{figure}
\subsection{Turf as a reference component} The reference dimension is chosen based on the exploratory disassembling of components. The skewness-kurtosis plot using the stochastic representation for Dirichlet random vector $y$ helped to disintegrate its parts. Based on the lowest kurtosis and skewness indicators the turf component is chosen as the $c^*$ reference (Figure \ref{fig:repstoc}). In coral reef community structures turf has been part of the major benthic groups.
\subsection{Modeling coral reef composition} Conditional to the reference component the model described in Section \ref{sec:model} was fitted to quantify the effects by sites, on corals, fire coral,  sponge, bryozoans, others, cca, cyanobacteria, macroalgae and turf from the coast of Southern Bahia of Brazil.
The independent normal prior distributions for the parameters $\beta$ and $\phi$ were used as described in Section \ref{sec:inference}.

Samples of the posterior distributions based on a Markov chain Monte Carlo (MCMC) method were used to obtain samples from the posterior distribution using the \cite{StanManual2018} in the R Environment.

Three Markov chains with length of 10,000 each, starting from different starting points with a warm-up of 9,000 iterations were generated. Convergence was visually verified via MCMC chain trajectories. Metrics such as Rhat values and the estimation of the effective sample size $(n_{eff})$ were considered as well.

Between Alg1-pgr2 and Alg2-pgr2G described in Section \ref{sec:comp-time}, the best model implementation to contrast the impact of habitat for this shared information approach between these different datasets was selected based on the log-likelihood for the Widely Applicable Bayesian Information Criterion (WAIC) and deviance information criterion (DIC) \cite{Gelman2014}. The computational time was reduced by 40\% when modeled using Alg2-pgr2G instead of Alg1-pgr2.

\begin{figure}[!ht]
	\centering
	\includegraphics[width=15cm,height=6cm]{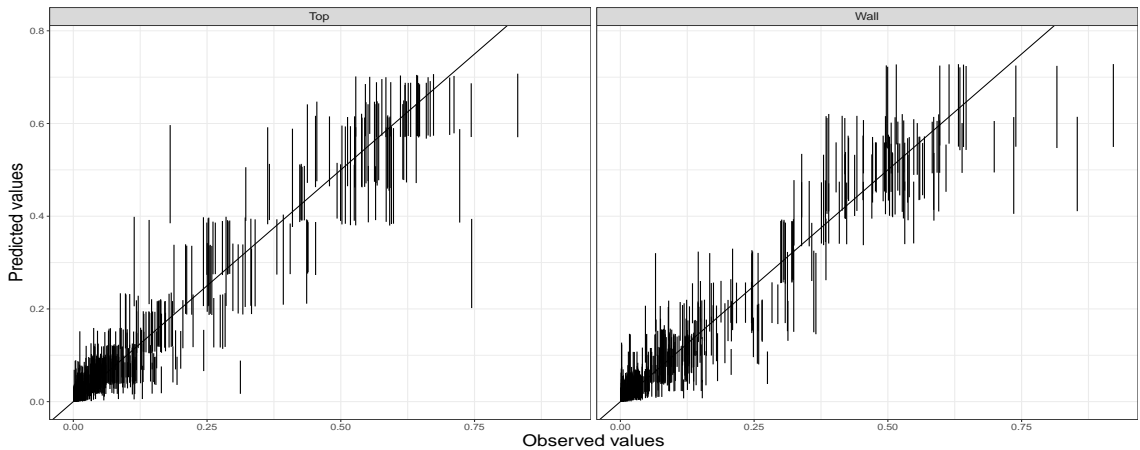}
	\caption{Observed values versus the posterior mean by habitat levels. Bars indicate the 95\% (black) and the 90\% (gray) credible intervals.}\label{fig:predobse}
\end{figure}

The model validation diagnostics included assessing the following: (a) goodness-of-fit by habitat in Table \ref{Tab:fitcompare}. (b) relationship between the model residuals and confirmation of the prediction capability.
\begin{table}[h!]
	\centering
	\begin{tabular*}{\textwidth}{@{}l@{\extracolsep{\fill}}l@{\extracolsep{\fill}}r@{\extracolsep{\fill}}r@{\extracolsep{\fill}}r@{\extracolsep{\fill}}r@{\extracolsep{\fill}}r@{\extracolsep{\fill}}r@{\extracolsep{\fill}}r@{\extracolsep{\fill}}r@{\extracolsep{\fill}}}
		\toprule
		\multirow{2}{*}{Model}&\multirow{2}{*}{Habitat}&\multicolumn{3}{c}{Fit}&&\multicolumn{4}{c}{Prediction}\\
		\cline{3-5}\cline{7-10}
		& & -2WAIC & pD & DIC& & aDist$^\mathcal{P}$ & Cover$^\mathcal{P}$ 95\% & rMSE$^\mathcal{P}$ & KL$^\mathcal{P}$ \\ 
		\midrule
		prg2G & Top & -283.008 & 19.930 & -267.189& & 1.906 & 0.935 & 0.023 & 0.114 \\ 
		prg2G & Wall & -294.330 & 23.936 & -274.095& & 1.918 & 0.923 & 0.022 & 0.101 \\
		prg2 & Top & -283.041 & 24.746 & -260.238& & 1.977 & 0.929 & 0.024 & 0.124 \\ 
		prg2 & Wall & -293.298 & 27.617 & -267.874& & 1.977 & 0.917 & 0.024 & 0.110 \\
		\bottomrule
	\end{tabular*}\label{Tab:fitcompare}
	\caption{Summary of the comparison criteria from fitted values under the proposed model against the observed value by sites. Widely Applicable Bayesian Information Criterion (WAIC), Deviance Information Criterion (DIC), aDist: Aitchison distance,
 Cover 95\%: 95\% coverage, 
		rMSE: the root of the mean-squared error,
		KL: Kullback Leibler divergence loss.}
\end{table}
The small values of the sDist, rMSE and KL metrics indicate good fits and predictions. Figure \ref{fig:predobse} represents these results in graphical form.
Figure \ref{fig:predobse} shows comparisons between credible intervals from predicted distribution against observed values. The exceptions are the few higher observed values which the model overestimated.
\subsection{The $\beta$ and $\phi$ effects on reef composition} 
\begin{figure}[!ht]
	\centering
	\includegraphics[width=15cm,height=6cm]{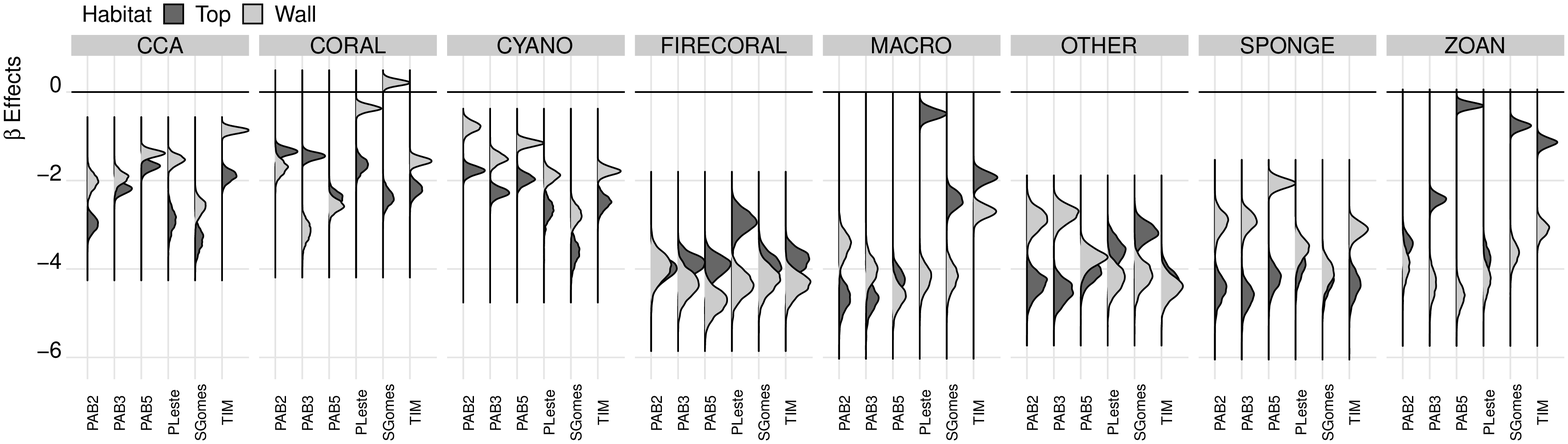}
	\caption{Density of the marginal posterior of the $\beta$ effects for each of the nine components by site and habitat}\label{fig:betaarticle}
\end{figure}

Figure \ref{fig:betaarticle} shows density of the marginal posterior of the $\beta$ effects for the two habitats. These results validate the original biologists’ hypothesis, that is, differences can be seen in various components and sites by habitat.
All components evidenced significant $\beta$ effects since the zero value is not contained in the credible interval. The effect size of PAB2 on the top habitat on the log-ratio scale between coral and turf functional groups decreased by 26\%. This means that the abundance of coral decreases in the PAB2 site. On the other hand, the wall habitat in Sebastião Gomes (SGomes) for the CORAL component had the only positive effect. The macroalgae (macro) component has the highest effect for PLeste-top. Effects related to cyano between habitats differ notably. This also happens with the Zoanthidea (zoan) component except on the PLeste site. Other interpretations can be drawn but this is outside the scope of this work.
\begin{figure}[!ht]
	\centering
	\includegraphics[width=1\linewidth]{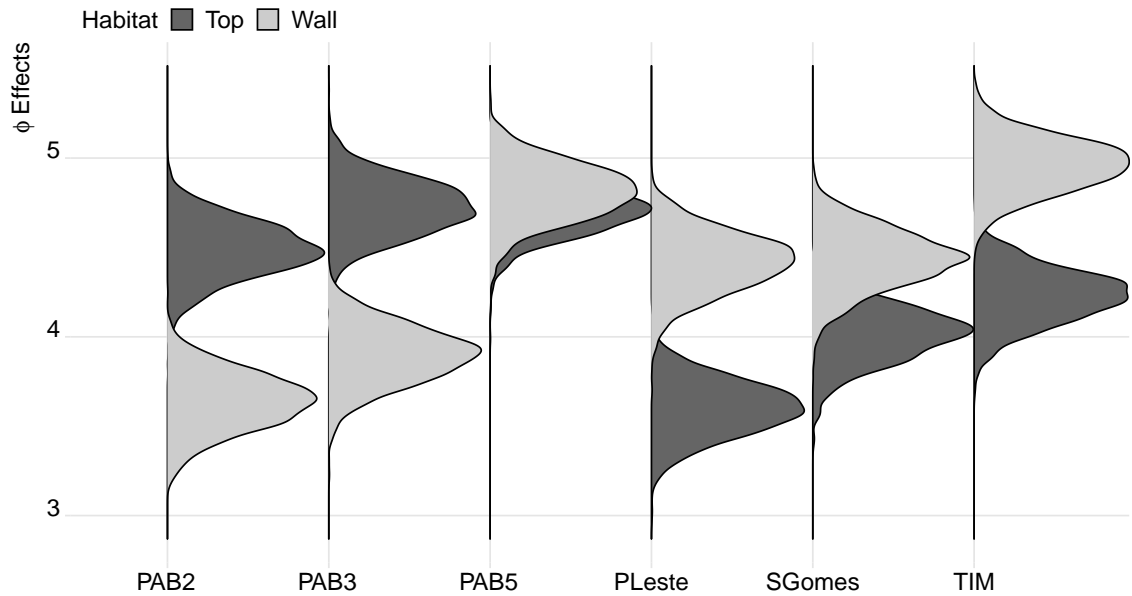}
	\caption{Variability by site: posterior density of the precision $\phi$ parameter by site and habitat level.}
	\label{fig:thetaarticle}
\end{figure}
The marginal posterior distributions of the parameters $\phi$ in Figure \ref{fig:thetaarticle} indicate that sites near the coast (inshore), PLeste, SGomes and TIM, have higher precision for the wall habitat than the top; the opposite occurs for the offshore sites (PAB2, PAB3). The exception is for the PAB5 $\phi$ effects, as they remain equivalent in both habitats. The posterior distribution for $\phi$ describes the patterns related to heterogeneity in the benthic process indicating the importance of the localization and habitat of each component. Note that this performance changes when different habitats are compared. 
These results show evidence of overdispersion and the benefit of using an appropriate model built to quantify effects, detect patterns and interpret this kind of data in the original scales.
\section{Conclusion and Future work}\label{sec:conclusion}
Variability based on composition information over different dynamic locations or datasets was studied in this paper. In the context of compositional data regression analysis, two contributions were presented. Regarding model identifiability, we presented a method to choose a component as a reference. The procedure consists of disassembling the composition using a sequence of independent gamma variables through a stochastic representation for the Dirichlet random vector. Then, an objective criterion based on the skewness and kurtosis metrics for each component was explored. This objective strategy based on properties related to the Dirichlet distribution supports choosing one component as a reference and is an alternative to the arbitrary selection of using component 1 or $C$. Despite yielding the same joint distribution, this choice can give a better interpretation of the results. Note that this specific choice is not unique or exclusive. Another procedure can be developed for other stochastic representations for Dirichlet random vectors and other characteristics can be required for the model construction. One other possibility is a beta representation.

The second contribution is based on building a hierarchical structure to integrate these multiple CoDa sets. This contribution combines Maier's model and the multilevel structure from \cite{Gelman-Hill2006}. This borrowed strength among the multiple CoDa sets (sites) induces a flexible hierarchical structure. Its strengths can be useful in many settings. For example, the proposed model allows us to share information on both the expected value and the precision. Both of these components can be structured to contain hierarchical parts (linear predictor). These advantages allow us to analyze the mean and precision components separately, which can be studied further. Additionally, the model formulation takes care to preserve the scale of the compositional response variables along with the statistical properties of multivariate data. Note that this kind of hierarchy differs from the proposal of \cite{Brewer2005} because it works with the shared information and they work via a distributions mixture.

The proposed model whose inference procedure was done with the Bayesian approach and estimation was carried out through MCMC methods via the Stan software \citep{StanManual2018} was validated under two perspectives. The analysis of simulated compositional data helps in the evaluation of the parameters estimation and prediction performance. The simulation scenarios were chosen considering the relationship between entropy and precision which allowed us to make the process inference with different levels of information since the entropy is implicitly determined.
This alternative parametrization incorporates crucial information related to the precision of the studied phenomenon.  

Concepts like skewness and kurtosis were explored to describe useful properties, like studying the tails to understand the presence or lack of outliers. This work does not combine components or categories. It provides a new rethinking about how to perform the alternative parametrization considering an objective choice of the reference component besides quantifying the gain on interpretation and results.

Following the original motivation, the composition data of a benthic coral reef community in the Abrolhos bank was studied. The model formulation takes care to preserve the compositional response variables along with the statistical properties of multivariate data used to estimate the relative proportions of this reef composition. In the marine ecological process, modeling the coral reef's dynamics provide an important initiative to understand the ecosystem for life underwater. The interpretation of these results on the biological area has the potential to contribute to the future of the Abrolhos area.

In the context of modeling multiple datasets, natural extensions and specific analyses including multiple covariates such as environmental variables can be incorporated to obtain a more realistic model. In an ecology context, this is not necessarily trivial because the mixture of different ecological sources of information introduces new different sources of uncertainty as well. Extensions with time-dependent effects must include special care to consider the original biologist’s experimental design in the model building. Furthermore, issues such as the inclusion of zeros or ones in the composition can be studied as well since this is not covered by the Dirichlet distribution.
\section*{Acknowledgements}
This work is supported by National Funds by FCT - Portuguese Foundation for Science and Technology, under the project UIDB/04033/2020.
Fieldwork and image processing were carried out by the team of the Marine Biodiversity and Conservation Laboratory of the Federal University of Rio de Janeiro. The author gratefully acknowledges the financial support of The Fundação Espírito Santense de Tecnologia, FEST. PM was funded by a scholarship from the Rio de Janeiro State Research Support Foundation (FAPERJ - E-26/200.016/2021 grant) also.

\end{document}